# ION CHAMBER ARRAYS FOR THE NUMI BEAM AT FERMILAB*

D.Indurthy, Ž.Pavlović, R.Zwaska, R.Keisler, S.Mendoza, S.Kopp,[#] M.Proga,
University of Texas, Austin, TX 78712, U.S.A.
D.Harris, A.Marchionni, J.Morfin, Fermilab, Batavia, IL 60510, U.S.A.
A.Erwin, H.Ping, C.Velissaris, University of Wisconsin, Madison, WI 53706, U.S.A.
M.Bishai, M.Diwan, B.Viren, Brookhaven National Lab, Upton, Long Island, NY 11973, U.S.A.
D.Naples, D.Northacker, J.McDonald, University of Pittsburgh, Pittsburgh, PA 15260, U.S.A.

*Abstract*

The Neutrinos at the Main Injector (NuMI) beamline will deliver an intense $\nu_\mu$ beam by focusing a beam of mesons into a long evacuated decay volume. We have built 4 arrays of ionization chambers to monitor the $\nu$ beam direction and quality. The arrays are located at 4 stations downstream of the decay volume, and measure the remnant hadron beam and tertiary muons produced along with neutrinos in meson decays.

## INTRODUCTION

The NuMI beamline [1] at Fermilab will deliver an intense $\nu_\mu$ beam to the MINOS detector in the Soudan Laboratory in Minnesota. Additional experiments are foreseen. The proton beam from the 120 GeV Main Injector is fast-extracted in 8.56 μsec spills onto the NuMI pion production target. The beam line is designed to accept $4\times10^{13}$ protons/spill. After the graphite target, two toroidal magnets called "horns" sign-select and focus forward the secondary $\pi$'s and $K$'s into a 675 m evacuated volume, where they may decay to muons and neutrinos. The horns and target may be positioned so as to produce a variety of neutrino beam energies [2]: moving the target further upstream of the horn produces a stiffer neutrino energy spectrum due to the nearly-constant transverse momentum of particles emerging from the target.

The secondary and tertiary beam monitoring system is shown in Figure 1. It monitors the integrity of the NuMI target and of the horns which focus the secondary meson beam. This monitoring is accomplished by measuring the lateral profile and intensity of the remnant hadron beam reaching the end of the decay tunnel of the muon beam penetrating the absorber and the downstream rock. Because muons are produced by the same pion decays as the neutrinos, the muon beam provides a good measure of the focusing quality of the neutrino beam.

## CHAMBER DESIGN

In ion chambers, slow-moving positive ions will screen the applied electric field of the chamber, increasing the drift time across the gas volume and the probability for charge recombination in the gas. Such loss of charge manifests itself as a non-linearity in ion chamber response *vs* the incident particle flux. Beam tests of our chambers under He gas flow indicate that the signal from He-gas chambers is linear up to intensities 40 times that expected at the NuMI monitors [3,4].

The hadron monitor is an array of 49 chambers mounted in a single Aluminum vessel (see Figure 2). The ion chambers are parallel plates made from ceramic wafers with Ag-Pt electrodes. The electrode separation is 1.0 mm. The vessel is sealed with an Pb-Sn wire, whose melting point of 200°C is well above the 50-60°C temperature at which the monitor will operate. The design maximizes use of Aluminum over stainless steel in order to reduce the presence of long-lived radionuclides in the detector: with its current proportion of 54lbs/4lbs Aluminum/Stainless, residual activation will be 58Rem/hr. The signal and HV are transmitted through 2 ceramic feedthroughs to custom-made cables constructed of aluminum core, insulated by a ceramic tube, shielded by an aluminum sheath. This cable transitions to a kapton-insulated cable. The detector slides into place through a slot in the absorber shielding (Figure 3).

The muon monitors are each 81 chamber arrays. The layout of 9×9 ion chambers is achieved by mounting 9 "tubes" onto a support structure vertically. Each tube has a tray of 9 ion chambers (3mm gap) with signals and HV routed to one end of the tube (see Figure 4). The endplates seal to the tubes using a Aluminum wire gasket. The signal and HV routing within the tube is accomplished with shielded kapton-insulated cable. Care was taken to minimize any exposed signal conductors to the gas volume, lest ionization in the surrounding gas collect on the conductors. Such "stray ionization" which occurs away from the ceramic chambers degrades the measurements of the muon beam's spatial profile. One detector plane is shown in Figure 5.

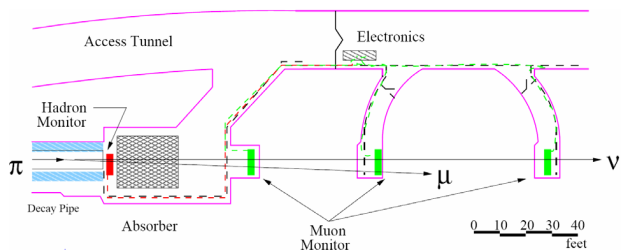

Figure 1: Layout of the secondary and tertiary beam monitors. The hadron monitor measures flux and spatial profiles of remnant hadrons. At three stations in the downstream rock, the muon monitors measure rates and spatial profiles of the muon beam. The NuMI target and horns are 675m upstream, to the left of this figure.

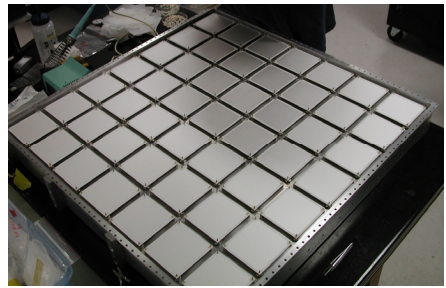

Figure 2: The hadron monitor during assembly showing the 7×7 array of ceramic ion chambers.

___________________________________________
*Work supported by U.S. DoE, contracts DE-FG03-93ER40757 and DE-AC02-76CH3000
[#]kopp@hep.utexas.edu

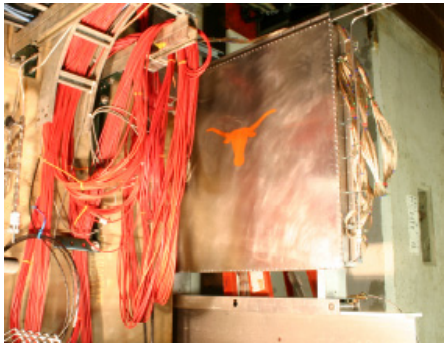

Figure 3: The hadron monitor during installation into the beam absorber The monitor is installed on rails through a 2m(H)×15cm(W) slot in the concrete shielding.

Tests were performed of the radiation damage to detector components at the University of Texas 1 MW fission reactor. Samples of chamber parts were exposed to $\sim 1.2 \times 10^{10}$ Rad.

## NUMI BEAM EXPERIENCE

The ion chamber arrays have been used to commission the NuMI beam in its first months of operation. A 1st run in December, 2004 successfully transported the primary beam into the target hall and to the beam absorber, followed by a second commissioning run in January, 2005, which delivered beam onto the target and furthermore focused the resulting secondaries into the decay volume. The hadron and muon monitors provided beam-based alignment of the target components and verification of the neutrino beam performance.

Figure 6 shows the Hadron and Muon Monitors' charge as a function of the proton intensity delivered to the target. The beam was set to the medium neutrino energy (ME) beam tune during this scan. The chambers show good linearity up to the $2.5 \times 10^{13}$ ppp maximum during this test. For reference, 1 pC in the muon chambers corresponds to $\sim 16,000$ (charged particles)/cm$^2$/spill.

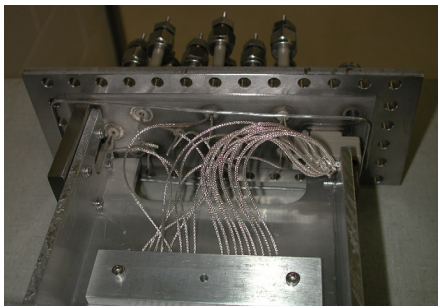

Figure 4: View of ion chamber tray being slid into rectangular muon tube.

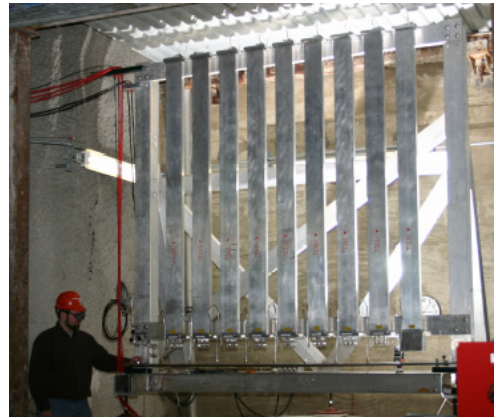

Figure 5: Array of 9 muon tubes in Alcove 1.

Figure 7 shows the muon chamber rate for each of the three alcoves as a function of the current in the focusing horns. As expected, turn-on of the horns results in improved focusing of pions and kaons produced in the target, resulting in greater muon (and presumably neutrino) flux at the end of the decay volume. The two plots in Figure 7 were taken in two separate neutrino runs, one in which the beam was set to a low neutrino energy tune, and one in which a high energy neutrino beam was produced. As can be seen in Figure 7, the higher energy neutrino beam does indeed produce greater muon flux in the downstream alcoves, which have higher expected thresholds due to their greater depth in the dolomite rock.

Figure 8 shows the total signal in the Hadron & Muon monitors as the proton beam is scanned across the target and protection baffle horizontally. Several features in these graphs permit use of this data to align the target and baffle with the primary beam. The hadron monitor shows two prominent spikes in response where the beam is transmitted through the gap between the target and inner aperture of the baffle (Figure 9). The asymmetry in the two peaks indicates that the target is ~0.25mm left-of-center within the baffle. The peaks' locations indicate that the target+baffle system is ~1.0mm left-of-center. Finally, the abundance of muons in the 3 alcoves as beam strikes the upstream collimating baffle is consistent with the expected behavior that a further upstream target produces a higher energy neutrino beam. The locations of the edges in the muon data establish the baffle position.

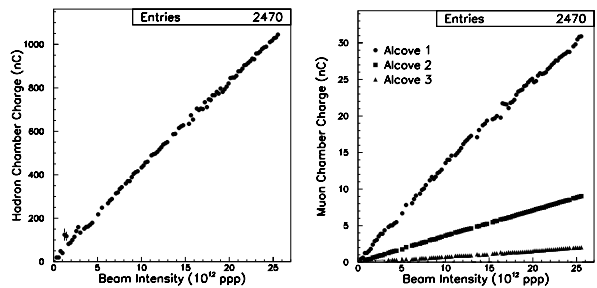

Figure 6: Charge in the hadron monitor and 3 muon monitors as a function of the protons on target while the beam is in the ME neutrino energy configuration.

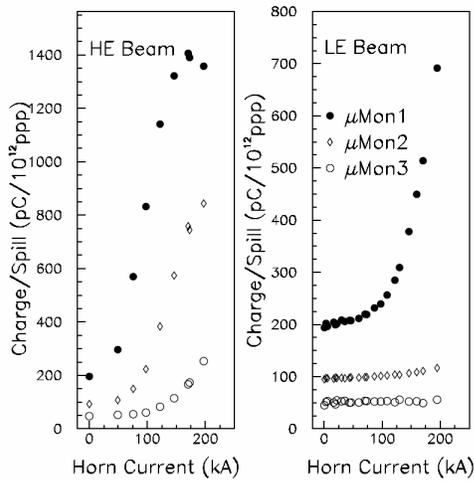

Figure 7: Total charge observed in the 3 muon monitor detectors as a function of the current in the focusing horns during a high energy (left) and low energy (right) neutrino run.

Figure 10 shows the muon beam centroid as the primary beam is scanned horizontally across the target. Of interest is the fact that, in the ME run, the centroid in alcoves 2 & 3 correlates with the proton beam, while in alcove 1 it anticorrelates with the proton beam. The correlation in alcoves 2 & 3 are expected because such muons originate from pions which pass through the field-free inner radius of the horns, thus track the proton beam directly, while the muons in alcove 1 are from softer, wider angle pions which are focused across the beam center line by the horns. In the HE run shown in Figure 10, a similar anticorrelation is seen in alcoves 1 & 2, since the HE beam focuses a larger pion momentum range through the horns.

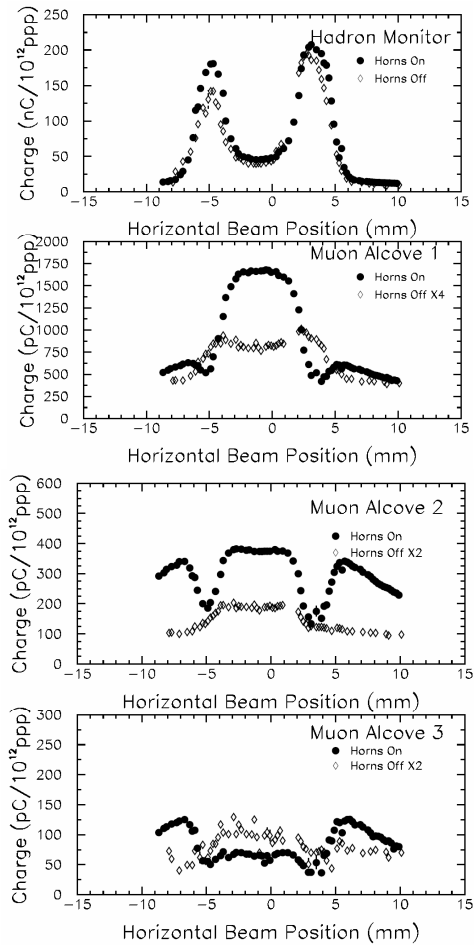

Figure 8: Response of the Hadron and Muon Monitors as the primary beam is scanned horizontally across the collimating baffle and the target. Shown are data acquired with the horns on & off during a ME neutrino run.

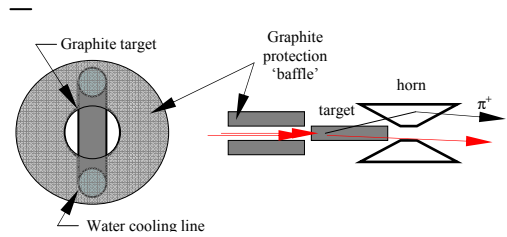

Figure 9: Schematic view end-on and side view of the protection baffle, target, and horns (not to scale).

## SUMMARY

We have built and commissioned an array of radiation-hard ionization chambers to monitor the NuMI neutrino beam at Fermilab. The system has contributed to alignment and commissioning of the beam line and is shown to be a effective monitor of beam quality.

## ACKNOWLEDGEMENTS

It is a pleasure to acknowledge the many contributions of our NuMI colleagues as well as from the MINOS collaboration in the successful startup of the NuMI beam.

## REFERENCES

[1] See S.Kopp, "The NuMI Beam at Fermilab," Fermilab-Conf-05-093-AD, *these proceedings*.
[2] M. Kostin *et al*, Fermilab note NUMI-B-0783 (2001).
[3] R.Zwaska *et al.*, *IEEE Trans.Nucl.Sci.* **50**,1129(2003).
[4] J.McDonald *et al.*, *Nucl.Instr.Meth*. **A496**, 293 (2003).

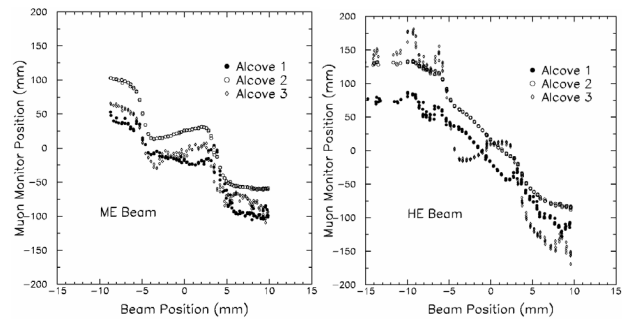

Figure 10: Horizontal muon beam centroid position as a function of the primary beam position on the target, for both a medium and a high energy neutrino run.